\documentclass[epjST]{svjour}
\usepackage{graphicx}
\usepackage{wrapfig}
\begin{document}
%Title of paper
\title{Multiple scroll wave chimera states}
\author{ Volodymyr Maistrenko\inst{1} \thanks{\email{maistren@nas.gov.ua}} 
\and  Oleksandr Sudakov\inst{1,2} 
\and Oleksiy Osiv\inst{1}  \and Yuri Maistrenko\inst{1,3}}
\institute{Scientific Center for Medical and Biotechnical Research, NAS of Ukraine, \\ Volodymyrska Str. 54, 01030 Kyiv, Ukraine \and  Taras Shevchenko National University of Kyiv, Volodymyrska Str. 60, 01030 Kyiv, Ukraine \and Institute of Mathematics, NAS of Ukraine, Tereshchenkivska Str. 3,  01030 Kyiv, Ukraine}
\abstract{
We report the appearance of three-dimensional (3D) multiheaded chimera states that display cascades of self-organized spatiotemporal patterns of coexisting coherence and incoherence.  We demonstrate that the number of incoherent chimera domains can grow additively under appropriate variations of the system parameters generating thereby head-adding cascades of the scroll wave chimeras.   The phenomenon is derived for the Kuramoto model of N identical phase oscillators placed in the unit 3D cube with periodic boundary conditions, parameters being the coupling radius $r$ and phase lag $\alpha$.  To obtain the multiheaded chimeras, we perform the so-called 'cloning  procedure' as follows:  choose  a sample single-headed 3D chimera state, make appropriate scale transformation, and put some number of copies of them into the unit cube. After, start numerical simulations  with slightly perturbed initial conditions and continue
them for a sufficiently long time to confirm or reject the state existence and stability.  It is found, by this a way,  that multiple scroll wave chimeras including those with incoherent rolls, Hopf links and trefoil knots admit this sort of multiheaded regeneration.  On the other hand, multiple 3D chimeras without spiral rotations, like coherent and incoherent balls, tubes, crosses, and layers appear to be unstable and are destroyed rather fast even for arbitrary small initial perturbations.}
\maketitle

\section {Introduction}
\hspace*{0.5cm} Chimera states represent one of the most fascinating discoveries of modern nonlinear science at the border of the network and chaos theories.  It has been found that networks of identical oscillators with non-local coupling can demonstrate robust co-existence of coherence and incoherence, such that a part of the network oscillators are synchronized but the others  exhibit desynchronized and often chaotic behavior.  First,  the chimera phenomenon was described in 2002 for the {\it one-dimensional} complex Ginzburg--Landau equation and its phase approximation, the Kuramoto model~\cite{kb2002}.  This paper  stimulated the study~\cite{as2004} two years later.  Both have opened a substantially new direction in the research of oscillatory  networks. See recent review papers on the topic~\cite{pa2015,es2016}. 

At about the same time,  this novel approach was extended to {\it two-dimensional} networks of oscillators.  Spiral waves with a
randomized core were identified for a class of three-component reaction-diffusion systems in the plane and for the two-dimensional Ginzburg--Landau equation as well as for the corresponding non-locally coupled Kuramoto model~\cite{ks2003,sk2004,kjm2004}.  They constitute a new class of chimera states, called the spiral wave chimeras~\cite{mls2010}.  This kind of spatio-temporal behavior is different from the oscillating 2D chimeras in which the coherent region only oscillates but does not spiral around the incoherence. The oscillating chimeras which are the natural counterparts of the 1D chimeras have been obtained in the form of stripes and spots, both coherent and incoherent~\cite{owyms2012,pa2013} and also twisted states~\cite{xkk2015}.
An interesting observation is that chimeras of both classes, oscillating and spiraling, emerge in opposite corners of the system parameter space and thus cannot co-exist~\cite{owyms2012}.

The first evidence of chimera states in {\it three-dimensions} was reported two years ago in ~\cite{msom2015} for the Kuramoto model of coupled phase oscillators in three-dimensional (3D) grid topology 
%\vspace*{-0.5cm} 
\begin{equation}
\dot{\varphi}_{ijk} = \omega+\frac{K}{N_{P}} \sum\limits_{(i^{\prime},j^{\prime},k^{\prime})\in B_{P}(i,j,k) }\sin(\varphi_{i^{\prime}j^{\prime}k^{\prime}} - \varphi_{ijk}- \alpha), 
%\ i,j,k = 1, ... , N,
\end{equation}
%\vspace*{-0.2cm}
%\hspace*{-0.55cm} 
where   $\varphi_{ijk}$ are phase variables, and  indexes $i,j,k$ are periodic mod $N$.   The coupling  is  assumed long-ranged and isotropic:  each oscillator $\varphi_{ijk}$ is coupled with equal strength $K$ to all its $N_{P}$ nearest neighbors $\varphi_{i^{\prime}j^{\prime}k^{\prime}}$  within  a ball of radius $P$, i.e., to those falling in the neighborhood 
%\vspace*{-0.5cm}
$$  B_{P}(i,j,k):=\{ (i^{\prime},j^{\prime},k^{\prime}){:} (i^{\prime}-i)^{2}+(j^{\prime}-j)^{2}+(k^{\prime}-k)^{2}\le P^{2}\},$$
%\vspace*{-0.5cm}
where the distances 
%$|i^{\prime}-i|,  |j^{\prime}-j|$, $|k^{\prime}-k|$ 
are calculated taking into account the periodic boundary conditions of the network.
The phase lag parameter $\alpha$ is assumed to belong to the attractive coupling range from $0$ to $\pi/2$.  
The second control parameter, coupling radius $r=P/N$ varies from $1/N$ (local coupling) to $0.5$ (close to global coupling). Without loss of generality, we put in Eq.(1) $\omega=0$ and $K=1$.

\begin{figure}[h]
\center{\resizebox{0.75\columnwidth}{!}{\includegraphics{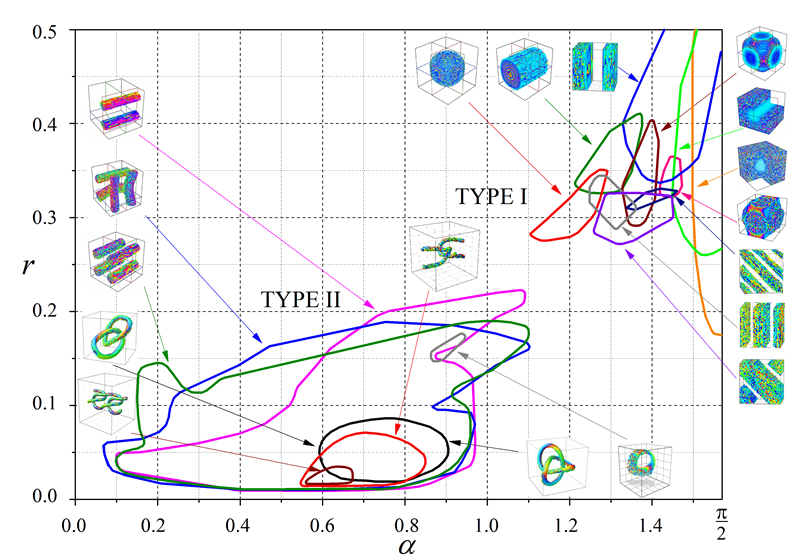}}} 
\caption{Parameter regions of 3D chimera states for Eq.(1). Regions for type I oscillating  and type II scroll wave chimeras appear in opposite corners of the parameter space.   
 Snapshots  of the states are shown in inserts. $r = P/N$. $N=100$.} 
 \label{fig:1}
\end{figure}

In~\cite{msom2015}, two principal families of 3D chimera states were obtained for Eq.(1): type I - oscillating chimeras, i.e., those without spiraling of the coherent region, and type II - spirally rotating chimeras, called {\it scroll wave chimeras}. Examples of the first class are coherent and incoherent balls, tubes, crosses, and layers in incoherent or coherent surrounding, respectively; the second class includes incoherent rolls of different modality and space disposition in a spiraling rotating coherent surrounding.   As it is illustrated in Fig. 1, parameter regions for chimeras of both classes (type I) and (type II)  do not intersect,  while there is a huge multistability inside each of the classes.  

Recently, two new kinds of the scroll wave chimeras, Hopf link and trefoil,  with linked and knotted incoherent regions (''swelling'' filaments) were detected in~\cite{ld2016}.  Our simulations confirm their existence in an ellipse-like parameter region which can be seen inside the type II chimera regions in Fig. 1 (delineated in black). Furthermore,  there exist in Eq.~(1)  scroll wave chimeras in the form of chains with one and two links. Parameter regions for one- and two-link chain chimeras are shown in Fig. 1 too (delineated in red and brown).  In the $R^3$-cube they look broken. However, they are indeed closed when considering them on the $T^3$-torus (which is topologically equivalent to the $R^3$-cube in the case of periodic boundary conditions).  
Note that all presented in Fig. 1 chimera patterns  are obtained with randomly chosen initial conditions. 

In the present paper, we study the appearance of multiheaded 3D chimera states built up on a base
of the single- and low-headed states exhibited in Fig. 1.   Similar to the 1D case~\cite{mvslm2014},  we design cascades of multiple 3D chimeras  with an increasing number of incoherent regions and obtain parameter regions for their existence.  To illuminate the multiheaded scroll wave  appearances, we apply the so-called  ''cloning procedure'' as follows:  glue a few copies of a chosen 3D chimera state, rescale them and stow them in the unit cube with periodic boundary conditions. Afterwards start simulation with the constructed multiheaded initial conditions perturbed slightly to prevent the symmetry capturing effect.

We show, with the use of massive numerical simulations,  that the cloning procedure perfectly works for the type II, i.e. scroll wave chimeras including rolls, Hopf links, and trefoils. The calculations were performed, as a rule, up to $t=10^{4}$ time units which corresponds to approximately 100 periods of the spiral rotations in the patterns. 

On the other hand, it fails for for the type I oscillating chimeras, as well as  chains. They disappear in the processes of simulation as soon as the symmetry imposed is violated.  Cascades of the roll-type scroll wave chimeras  with even numbers of  incoherent rolls  are constructed in Ch. 2.  Hopf link and trefoil cascades are obtained in Ch.3.  ''Hybrid chimeras'' including different combinations of trefoil, Hopf links and parallel rolls are illustrated in Ch. 4.   Examples of the multiheaded scroll wave dynamics are demonstrated by videos at 
http://chimera3d.biomed.kiev.ua/multiheaded.

Numerical simulations were performed on the base of Runge-Kutta solver DOPRI5 that was integrated into
the software for large nonlinear dynamical networks~\cite{sls2011}, allowing for parallelized simulations
with different sets of parameters and initial conditions. The simulations were performed at the computer cluster ''CHIMERA'', http://nll.biomed.kiev.ua/cluster, and the Ukrainian Grid Infrastructure
providing distributed cluster resources and the parallel software~\cite{zssb2007}.

\section {Cascades of scroll wave chimeras with multiple incoherent rolls}

\begin{figure}[ht!]
 \center{\includegraphics[width=0.27\linewidth]{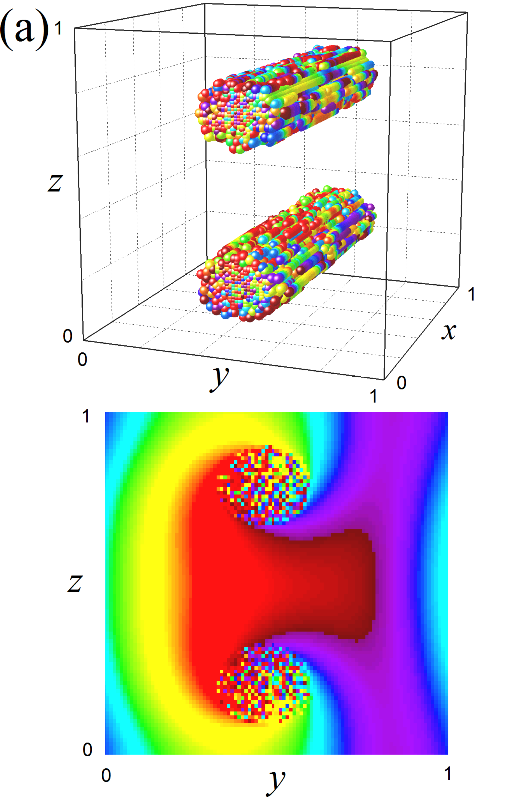} 
  \includegraphics[width=0.27\linewidth]{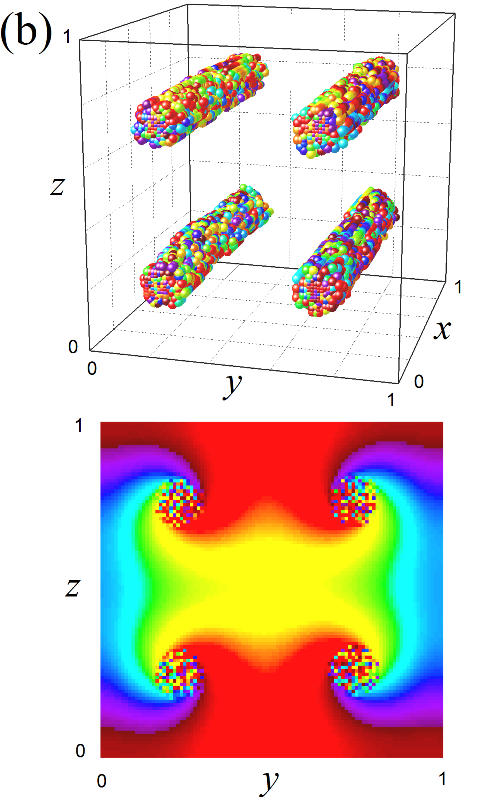}  
  \includegraphics[width=0.295\linewidth]{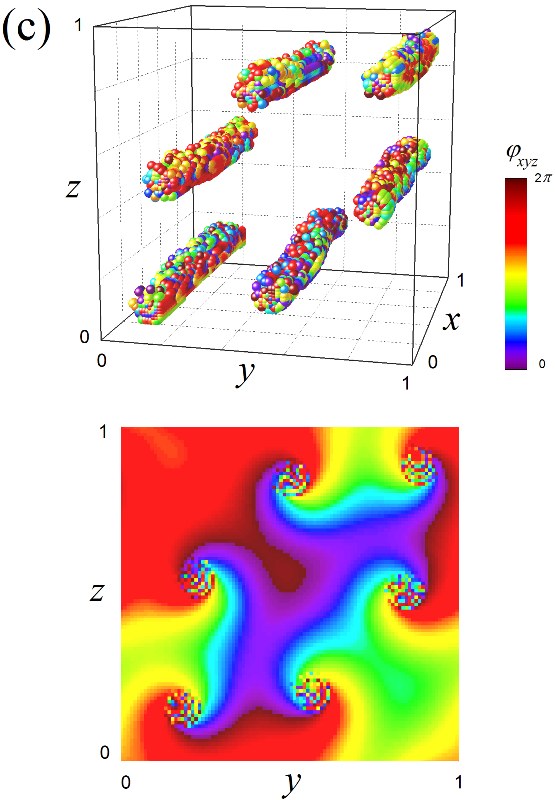}} 
 \center{\includegraphics[width=0.27\linewidth]{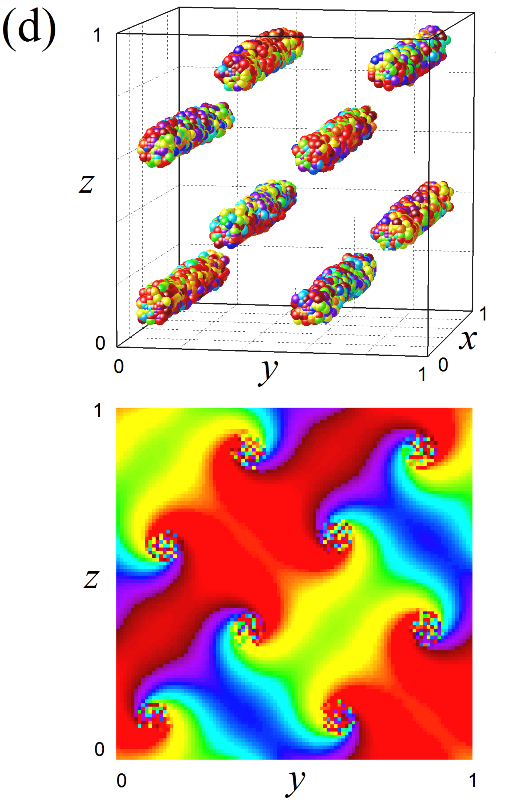} 
 \includegraphics[width=0.265\linewidth]{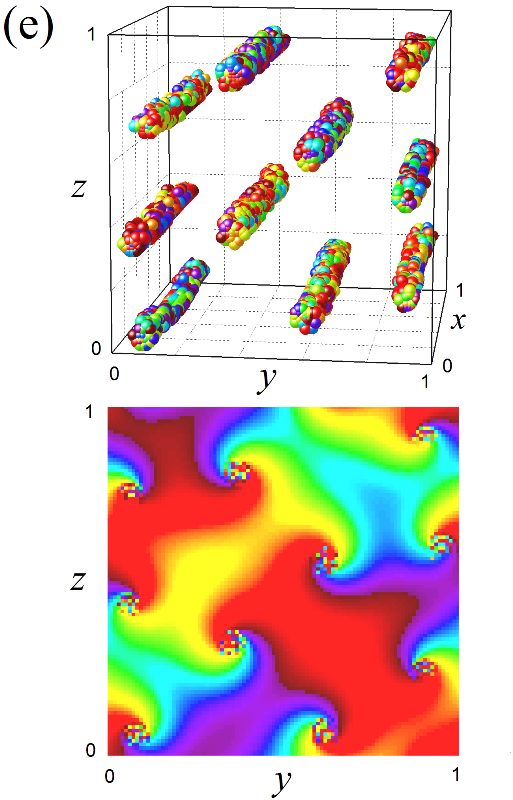} 
  \includegraphics[width=0.295\linewidth]{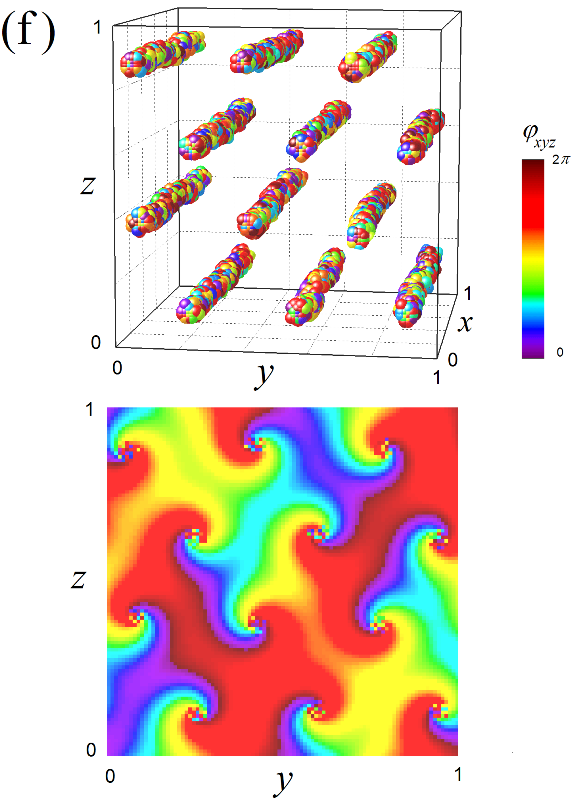}} 
 \center {\includegraphics[width=0.265\linewidth]{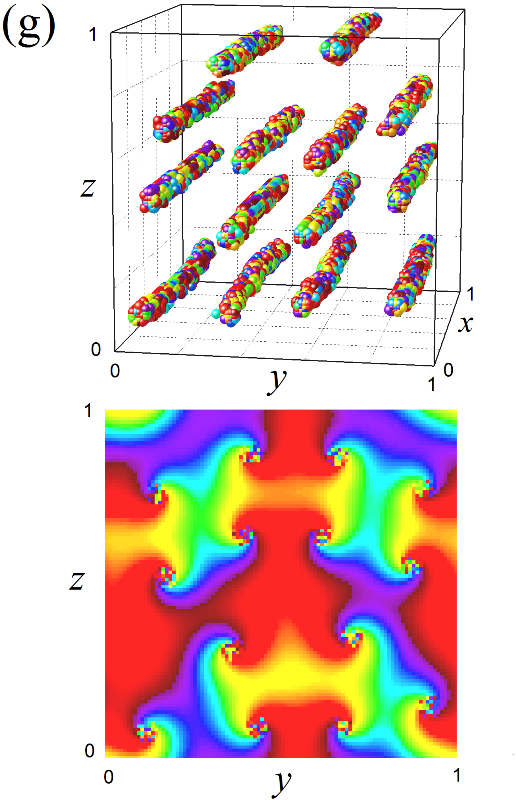}  
 \includegraphics[width=0.265\linewidth]{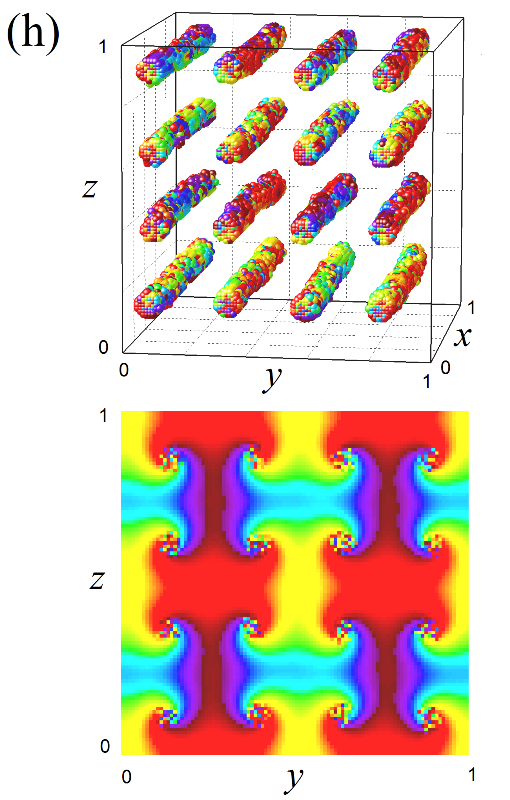} 
 \includegraphics[width=0.295\linewidth]{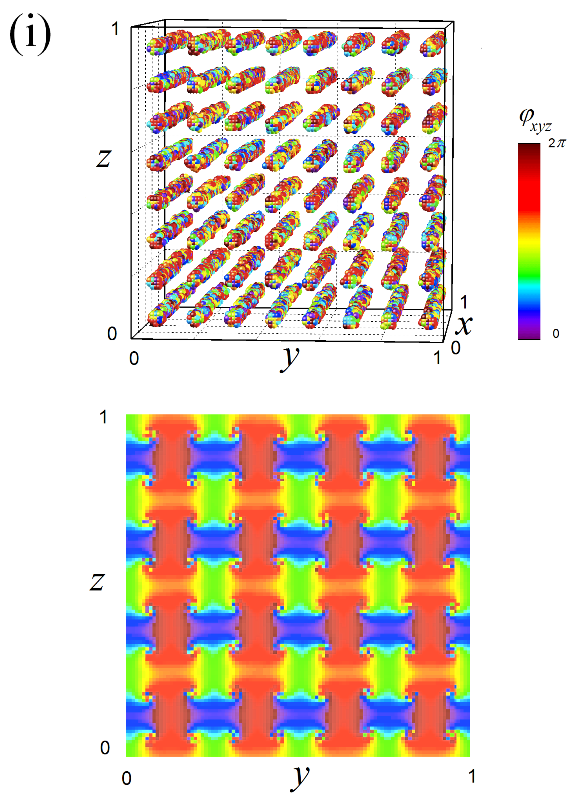}}
 \caption{Cascade of multiheaded scroll wave chimera states with parallel rolls.  3D screenshots and respective cross-sections are shown for: (a) - two rolls ($\alpha=0.8, r=0.165$), (b) - four rolls ($\alpha=0.7, r=0.12$), (c) - 6 rolls ($\alpha=0.7, r=0.09$), (d) - 8 rolls ($\alpha=0.7, r=0.08$),
(e) - 10 rolls ($\alpha=0.64, r=0.07$), (f) - 12 rolls ($\alpha=0.63, r=0.05$), (g) - 14 rolls ($\alpha=0.64, r=0.056$), (h) - 16 rolls ($\alpha=0.6, r=0.06$), $N=100$; (i) - 64 rolls ($\alpha=0.4, r=0.04$). $N=200$. 
 Coordinates  $x=i/N,  y=j/N,  z=k/N$. } 
  \label{fig:2}
\end{figure}

\hspace*{0.5cm} Scroll waves with parallel incoherent rolls represent one of the characteristic examples of the chimera states in three-dimensions. In~\cite{msom2015}, 2- and 4-rolled chimeras of this type were obtained, they exist in wide regions of the ($\alpha,r$)-parameter space shown in Fig. 1.  Due to the periodic boundary conditions they can be considered as scroll rings with incoherent cores (''swelling'' filaments) on the $T^3$-torus.  The microscopic dynamics inside the chimera rolls is chaotic while, in the large-scale, the rolls themselves are practically stationary, i.e. not moving  in a significant manner. This can be seen in the supplementary videos of ~\cite{msom2015},  also at http://chimera3d.biomed.kiev.ua/high-resolution (files fig5(a)hq-video.mkv, fig5(b)hq-video.mkv); as well as for 16, 64 parallel and 16 crossed rolls 
 in http://chimera3d.biomed.kiev.ua/multiheaded/rolls/.

In Fig. \ref{fig:2}, a cascade of  scroll wave chimeras with pair-multiple incoherent parallel  rolls is presented. The number of rolls  increases additively from 2 [Fig.\ref{fig:2}(a)] to 
16 [Fig.\ref{fig:2}(h)].  In addition, a 64-rolled chimera is shown in [Fig.\ref{fig:2}(i)].  The chimera rolls in Fig.\ref{fig:2} are symmetrically located in the unit cube.  However, the large-scale symmetry can be violated when other initial conditions are chosen.  Note also that microscopic chaotic dynamics inside the rolls differ for different  rolls of the same state, which is  illustrated by the cross-sections shown below the 3D plots in Fig.\ref{fig:2}. 

 \begin{figure}[ht!]
\center{\resizebox{0.53\columnwidth}{!}{\includegraphics{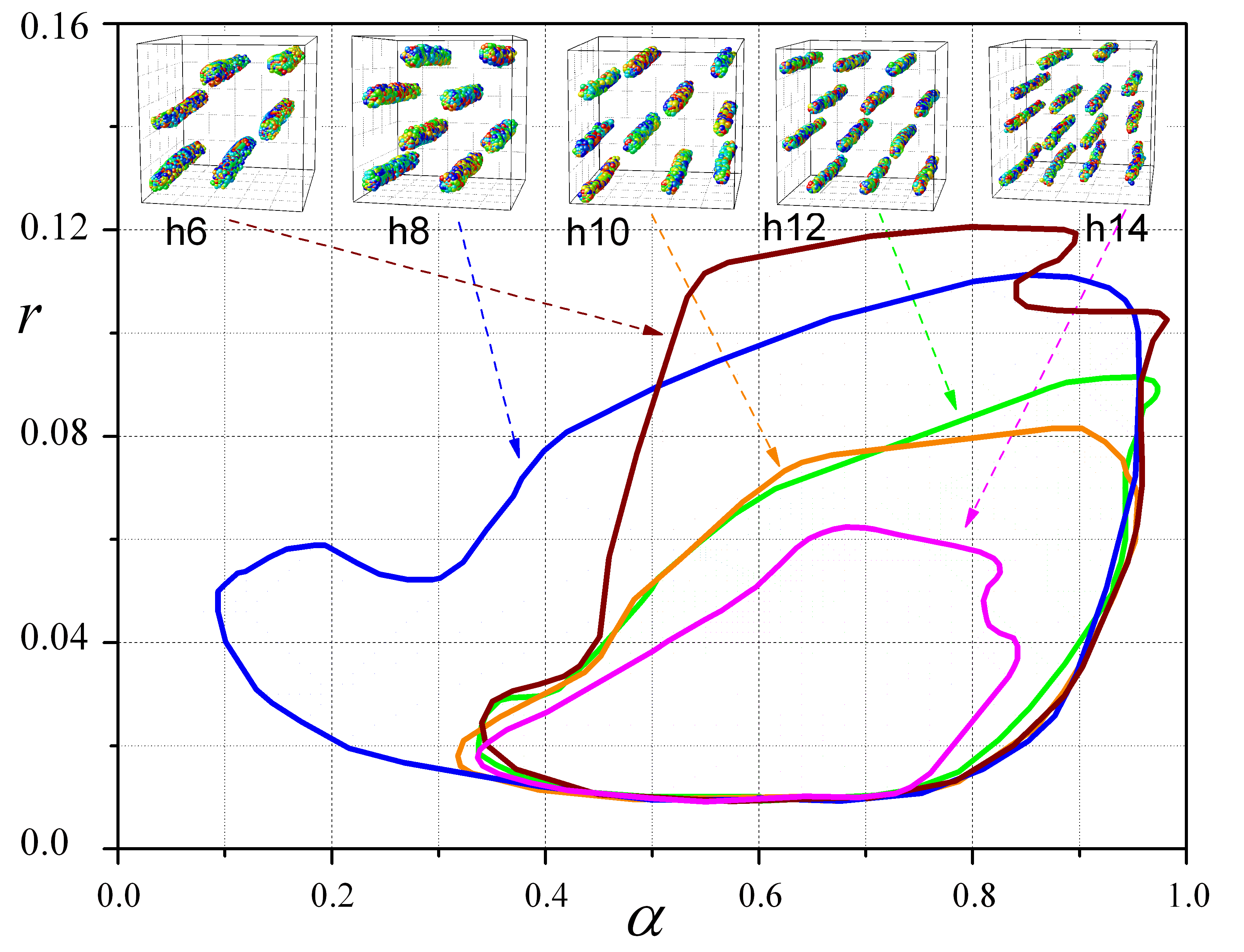}}}
\caption {Parameter regions for $h$-headed parallel rolls chimera states delineated by the color lines: 
blue - h6, brown - h8, orange - h10 , green - h12, magenta - h14 rolls. 
 $r=P/N, N=100$. Snapshots of respective chimera types are shown in inserts.}
  \label{fig:3}
\end{figure}

Regions for existence of the parallel rolled chimeras  in the $(\alpha,r)$-parameter plane are shown in Fig.\ref{fig:3}.  They lie at intermediate phase shifts  $\alpha$  between $0.1$ and $0.95$,  and at small coupling radius $r<0.12$ including the minimal possible $r=0.01$   when only 6 nearest neighbors are connected to each oscillator ($N=100$).  Thus, the multiple rolled chimeras exist in the model (1) not only for non-local but also for the local coupling scheme. Based on this, we assume that  such states should exist also in the limiting PDE case $N\rightarrow\infty, r\rightarrow 0$.  If so, the PDE obtained by this a way could be a rich source for multiple scroll waves (multiple scroll rings when written in the circular coordinates).  Our simulations confirm that chimeras illustrated in Fig. 2 are robust 3D patterns as they survive for long integrating times of  thousands of rotating periods.  Further study in this direction would be interesting from both theoretical and practical points of view;  е.g.,  in medicine as prospective models of spiral patterns formed on heart tissue during
ventricular tachycardia and fibrillation (see~\cite{pa2015,msom2015} and references therein).

To obtain the multi-rolled scroll wave chimeras we used the so-called 'cloning procedure' as follows. First, take some number of 2- or 4-rolled parallel chimeras (previously obtained in ~\cite{msom2015}). Rescale them in an appropriate way and fill the unit cube with them. Afterwards, start calculations with these specially prepared initial conditions slightly perturb to prevent the symmetry capturing effect. Doing so,  there is no guaranty that the resulting state will be of the form as assigned,  i.e. with the initially chosen number of rolls. We often had to repeat the procedure trying different variants of the number and type of initial 'chimera clones', as well as varying system parameters. Proceeding in such a way,  after some number of trials the desired chimera pattern was usually obtained. 

\begin{figure}[h]\sidecaption
\resizebox{0.65\columnwidth}{!}{\includegraphics{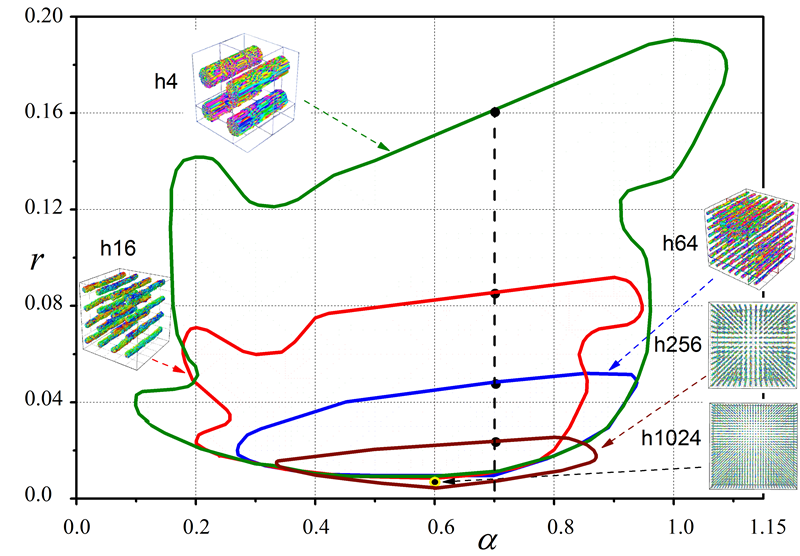}}
%\begin{figure}[ht!]
 % \center{\includegraphics[width=0.55\linewidth]{Fig4.png}}
\caption{Parameter regions for $h$-headed parallel rolled chimera states delineated by the color lines: olive - h4 ($N=50$), red - h16 ($N=100$), blue - h64 ($N=100$), brown - h256 ($N=200$) rolls. 
  $r = P/N$. Snapshots of respective 3D chimera types are shown in inserts.}
  \label{fig:4} 
\end{figure}

\begin{figure}[ht!]
 \center{\includegraphics[width=0.6\linewidth]{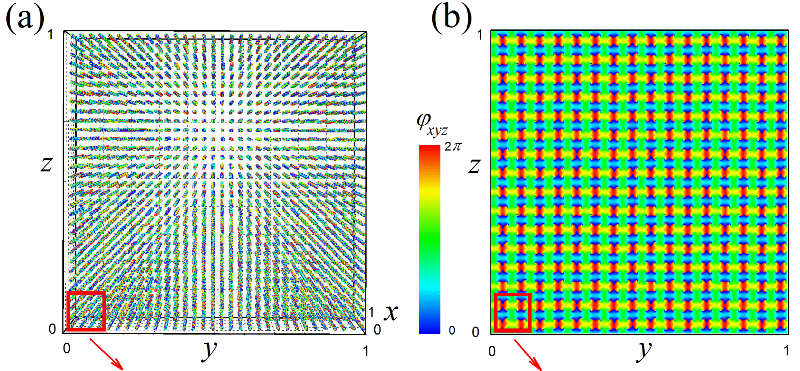}}
 \center{\includegraphics[width=0.62\linewidth]{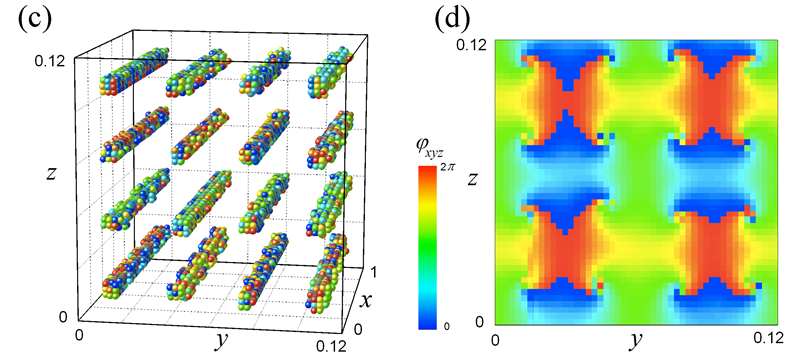}}
\caption{Example of 1024-headed parallel scroll wave chimera state (a), and its cross-section at $x=0.5$ (b) with enlarged windows (c,d) ($\alpha=0.6, r=0.0075, N=400$).}
  \label{fig:5} 
\end{figure}

Among other variants of the cloning procedure, there is one reliable approach always giving the chimera state we are looking for. This is in the case when just 8 identical  parallel chimeras are taken as samples and placed, after rescaling, in the unit cube. Then, a 4-multiple rolled chimera is that is stable with respect to perturbations is obtained. We have never seen its destruction even at very long simulations.  Therefore, we can successively repeat the multiple  chimera regeneration as long as computer power allows to process it.  As the system complexity grows exponentially, we were only able to produce $4^n$-rolled chimeras with $n=1,2,3,4$ and $5$.  Our largest example is the 1024-rolled chimera calculated for $N=400$, i.e. for  $N^3=64$ million  oscillators up to 1000 time units, illustrated in Fig.\ref{fig:5}.

Stability regions for 4-, 16-, 64- and 256-headed  parallel scroll wave chimera states
are presented in Fig.\ref{fig:4}; the parameter point for the 1024-headed chimera in Fig.\ref{fig:5} is also shown.  As it can be observed,  each next region in the cascade is twice thinner on the parameter $r$ compared to the previous one. 
E.g.,  at $\alpha=0.7$ the top border values of parameters $r$  of the stability regions decreases approximately as 0.16, 0.08,  0.04, 0.02. 
We assume that with more computational power,  the whole cascade  can be obtained  for the 4-multiple scroll wave chimeras with any $2^{2(n+1)}, n=1,2,3...$ number of heads. Note that this cloning procedure can be also applied successfully for 4 crossed rolls chimera.

\section {Hopf link and trefoil chimera states}
\label{sec:3}

\begin{figure}[ht!]
  \center{\includegraphics[width=1\linewidth]{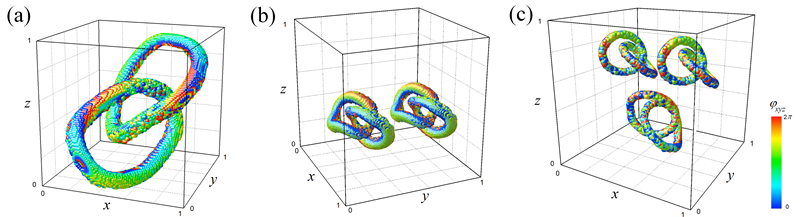}}
  \center{\includegraphics[width=1\linewidth]{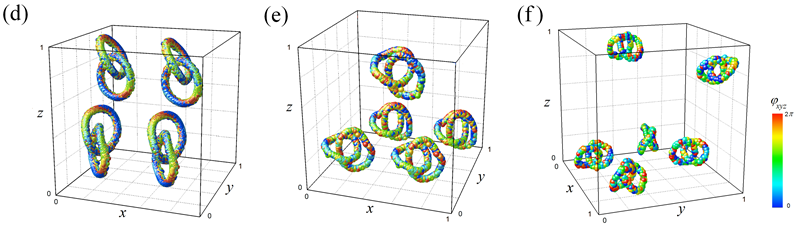}}
  \center{\includegraphics[width=0.63\linewidth]{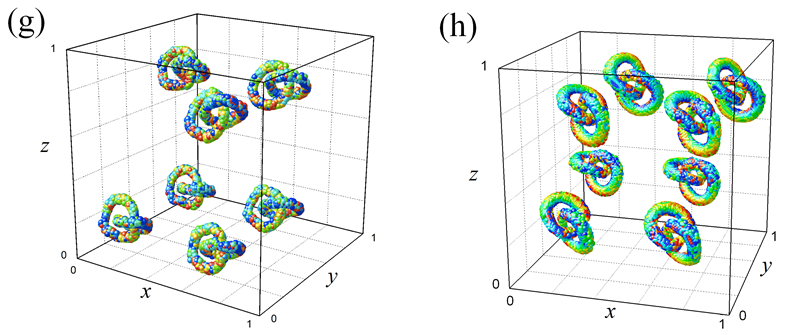}
\includegraphics[width=0.35\linewidth]{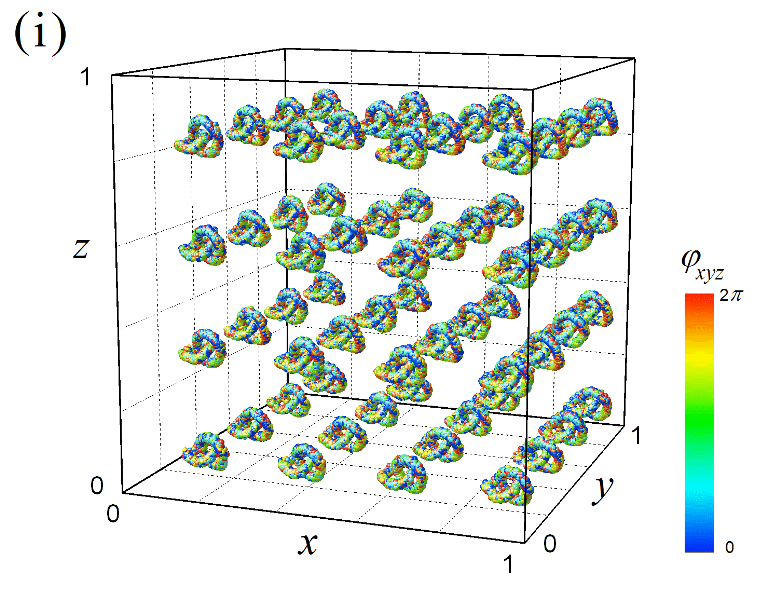}}
\caption{Cascade of multiple Hopf link chimera states:  (a) - 1-headed ($\alpha=0.76, r=0.083, N=100$), (b) - 2-headed ($\alpha=0.61, r=0.0215 $), (c) - 3-headed ($\alpha=0.61, r=0.027$), (d) - 4-headed ($\alpha=0.68, r=0.03$),  (e) - 5-headed ($\alpha=0.68, r=0.018$), (f) - 6-headed ($\alpha=0.7, r=0.017$),  (g) - 7-headed ($\alpha=0.74, r=0.02$), (h) - 8-headed ($\alpha=0.84, r=0.03$). $N=200$; (i) - 64-headed ($\alpha=0.72, r=0.01$), $N=400$.} 
\label{fig:6}
\end{figure}

 Hopf link and trefoil chimera states represent 3D scroll waves with linked and knotted  filaments. Due to the non-local coupling, the filaments are not singular (lines) as in standard  scroll waves but ''swelled'' proportionally to the radius of coupling. Moreover, they are filled by oscillators with unsynchronized, chaotic  behavior.  The Kuramoto model, Hopf link and trefoil chimeras were fist reported in~\cite{ld2016}, see also~\cite{fm2000} where bifurcation transitions between this kind of 3D patterns are studied for a different model with local coupling.

\subsection{Cascades of multiple Hopf links and trefoils}

\begin{figure}[ht!]
  \center{\includegraphics[width=1\linewidth]{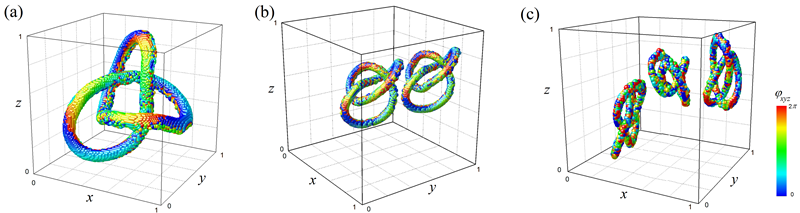}}
\center{\includegraphics[width=1\linewidth]{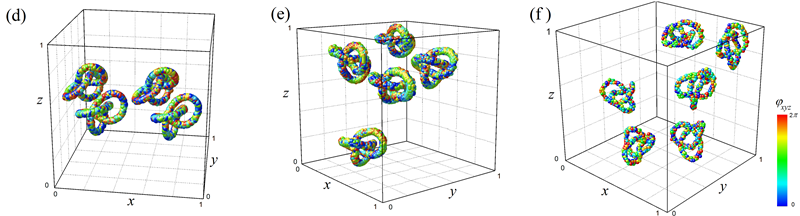}}
  \center{\includegraphics[width=0.63\linewidth]{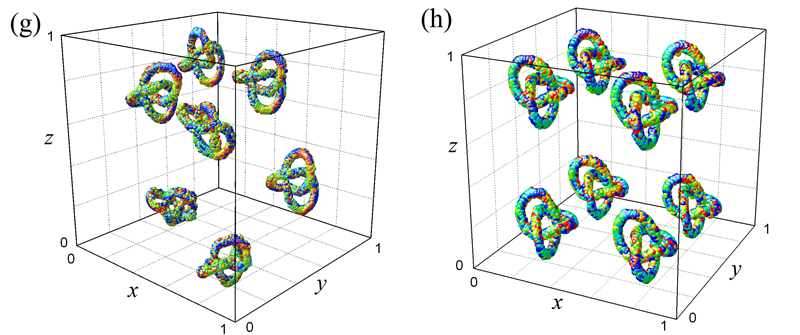}
\includegraphics[width=0.35\linewidth]{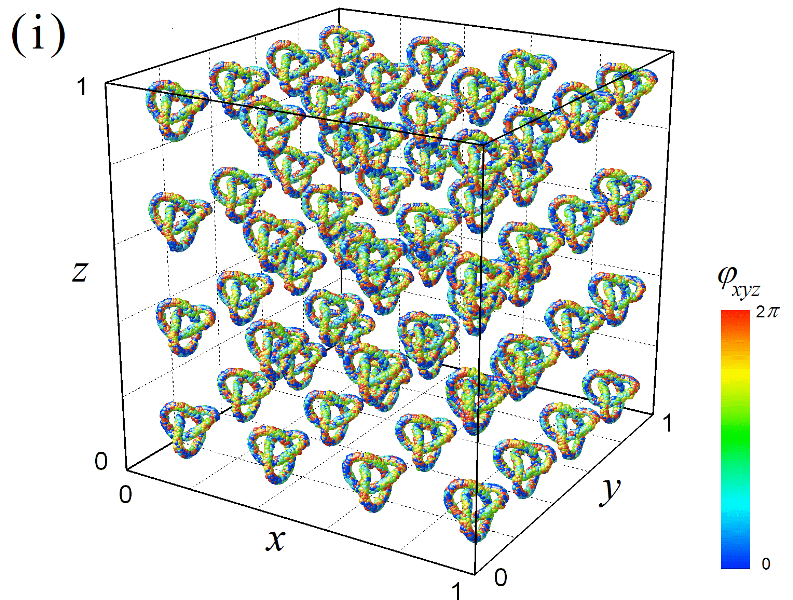}}
\caption{Cascade of multiple trefoil chimeras states:  (a) - 1-headed ($\alpha=0.68, r=0.07, N=100 $); (b) - 2-headed ($\alpha=0.6, r=0.023 $),  (c) - 3-headed ($\alpha=0.68, r=0.02$), (d) - 4-headed ($\alpha=0.585, r=0.02$), (e) - 5-headed ($\alpha=0.672, r=0.02$), (f) - 6-headed ($\alpha=0.72, r=0.02$), (g) - 7-headed ($\alpha=0.78, r=0.02$), (h) - 8-headed ($\alpha=0.72, r=0.02$), $N=200$; (i) - 64-headed ($\alpha=0.72, r=0.01$), $N=400$.} 
  \label{fig:7}
\end{figure}

In this chapter we design cascades of Hopf link and trefoil chimera states with additively growing the number of samples.  
Figs.\ref{fig:6} and \ref{fig:7}  illustrate the $h$-multiple cascades of Hopf links and trefoils, respectively,  for $h=1,2,3,...,8$ and $64$. They are constructed by the cloning procedure using the the sampled initial conditions of identical Hopf links or trefoils, always with small perturbations to prevent the symmetry capturing.   Constructed Hopf links and trefoils are not stationary patterns (in contrast to the parallel scroll waves in the previous Chapter), this is illustrated by videos at
http://chimera3d.biomed.kiev.ua/multiheaded.

\begin{figure}[h]
  \center{\includegraphics[width=0.7\linewidth]{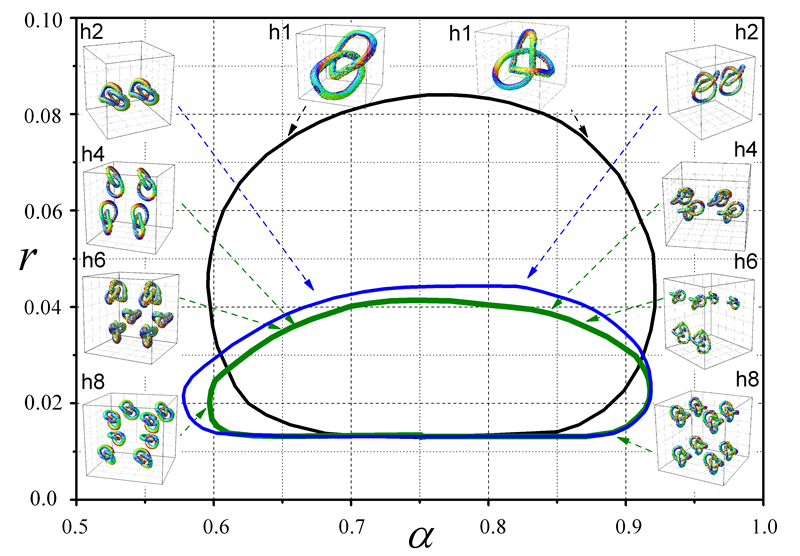}}
\caption{Even cascades of Hopf link and trefoils:  one-headed (delineated by black line), 
2-headed (delineated by blue line), 
4-headed,
6-headed and 
8-headed (delineated by green line).
$r =P/N, N=200$. Snapshots of the chimera states are shown in inserts.}
  \label{fig:8} 
\end{figure}

We find that  multiple chimera states of this kind exist in wide enough regions of the ($\alpha,r$)-parameter plane. The regions are heavily intersecting and do not shrink as $h$ increases. This is illustrated by the bifurcation diagrams in 
Fig. \ref{fig:8} for $N=200$.
To our surprise,  both parameter regions for single Hopf link and single trefoil coincide (delineated in black in Fig. 8). The same  occurs for the respective multiple patterns.  Indeed,  2-Hopf link and 2-trefoil states co-exist in the twice smaller region (delineated in blue).  Moreover, the states of higher multiplicity $h=3,4,...,8$,  are all found in the same slightly smaller inclusive region (delineated in green).   Therefore, parameter regions for the $h$-multiple states stabilize as $h$ increases 
and,  given our precision, they become indistinguishable beginning from $h=3$.

Regions for Hopf link and trefoil chimera states are located in the intermediate range of the phase lag parameter $\alpha$, approximately between 0.6 and 0.9, and for rather small values of the coupling radius $r=P/N<0.08$. The lower boundary of the regions is given by the value $r=0.012245...$ ($N=200$). This is the smallest value of $r$,  when the number of nearest neighbors  $\varphi_{i^{\prime}j^{\prime}k^{\prime}}$  coupled to each oscillator $\varphi_{ijk}$  is $N_{P_{min}}=81$. At smaller $r<0.012245$, the number of coupled oscillators drops abruptly to 59.  Our simulations confirm that both single and multiple Hopf links and trefoils do not survive with such low connectivity ($P=59$ or smaller) and are fast transforming into some other state. We conclude that Hopf links and trefoils arise in the Kuramoto model (1) only with non-local, prolonged coupling.  Local diffusive coupling is not enough to insure their stability, which is unlike to the roll-type chimeras surviving at local coupling, see Ch.2.

Multiple trefoil and Hopf link chimeras were obtained using the cloning procedure.  It perfectly works, however, only when just 8 smaller copies of a state are laid into the unit cube. 
The other combinations, e.g., when looking for 2h, or 3h states,   do not guarantee the desired result and require as a rule  additional efforts.  In many such cases the multi-compound structure appears to be unstable and is destroyed rather fast as simulations start. Then, we have to try again with different initial conditions and system parameters until the desired multi-headed Hopf link ot trefoil is eventually obtained.  
 
\vspace*{-0.5cm}
\subsection{Chain chimeras}

\hspace*{0.5cm} For the single-link chimera two more kind of linked scroll wave chimeras are given by single- and double-link chains, illustrated in Fig.9.  In the $R^3$-cube the chains are broken, but  they are indeed connected on the corresponding $T^3$-torus.   Parameter regions for the chain chimeras are presented in Fig. 10. As it can be seen, similar to the Hopf links and trefoils, they both arise in Eq. (1) at the intermediate values of the phase lag parameter $\alpha$ and for a rather small radius of coupling. For single-link chains the parameter $\alpha$ should be approximately between 0.55 and 0.85,  and $r$ be smaller than 0.075. 

\begin{figure}[ht!]
  \center{\includegraphics[width=0.8\linewidth]{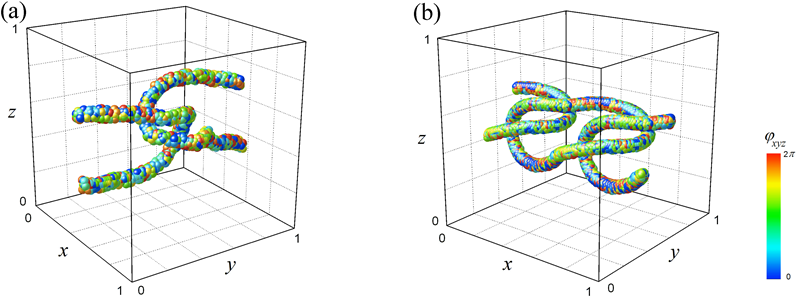}}
  \caption{Chain chimera states: (a) -  single-linked chain ($\alpha=0.7, r=0.041, N=100$), (b) - double-linked chain ($\alpha=0.61, r=0.033, N=200$).}
  \label{fig:9} 
\end{figure}

The lower boundary of the single-link chain chimera is given by the coupling radius value  $r=0.00707...$ when each oscillator in the network is coupled to $N_P=19$ its nearest-neighbors. At smaller $r$ the number $N_P$ of the couplers drops abruptly to $7$ only. The state becomes unstable and is rapidly destroyed in the simulations.  Parameter regions in Fig. 10 are obtained for Eq.(1) with $N=200$.  Interestingly,  the same $N_P=19$ lower bound on the number of coupled oscillators is also obtained for the chain chimera stability in Eq.(1) with  $N= 100$.

\begin{figure}[h]\sidecaption
\resizebox{0.63\columnwidth}{!}{\includegraphics{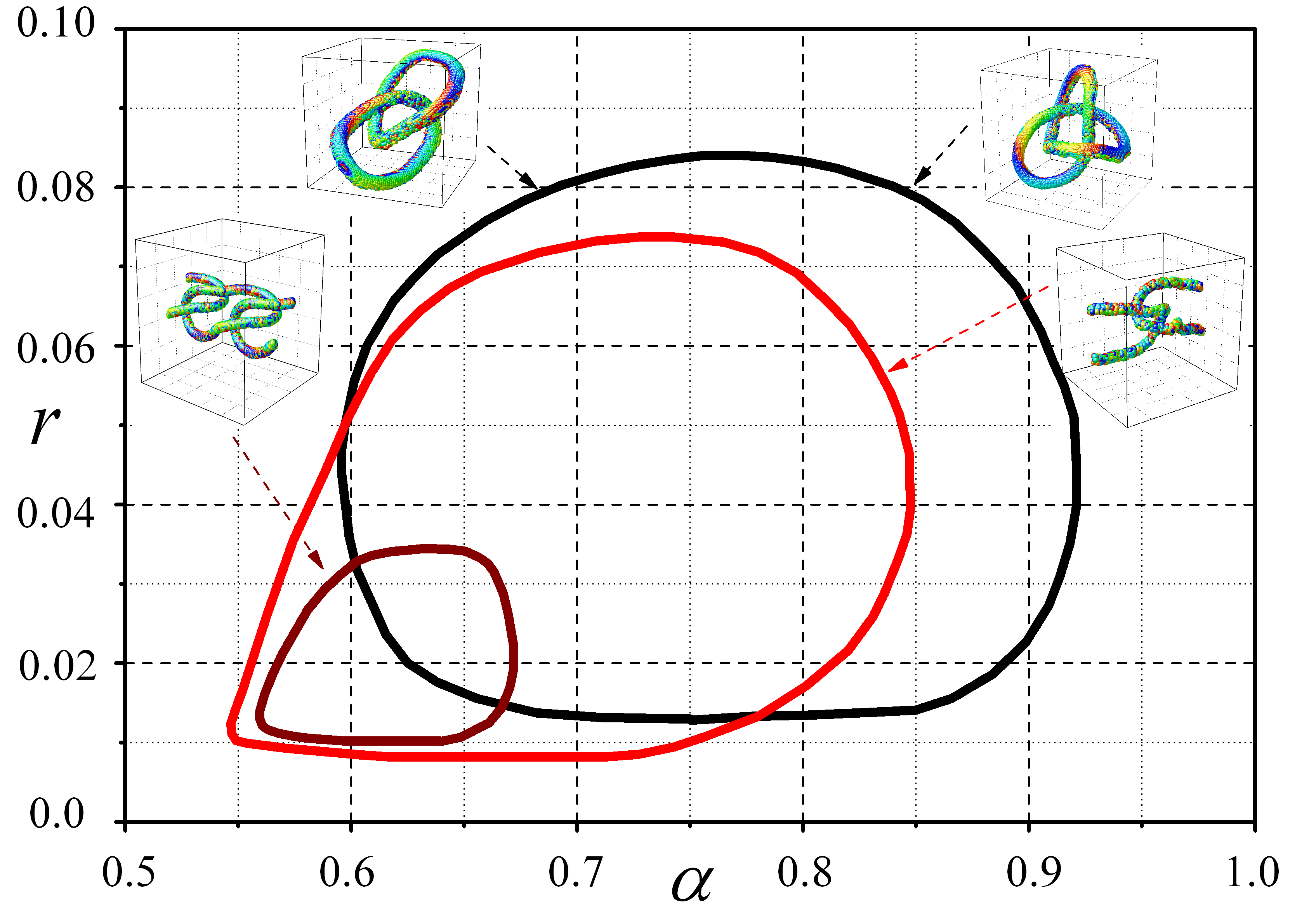}}
\caption{Parameter regions for single-link (delineated by red line) and double-link (delineated by brown line) chain chimera states. Region for  single Hopf link and trefoil are also shown (delineated by black line).
$r =P/N, N=200$. }
\label{fig:10}
\end{figure}

Note that case of $N_P=19$ couplers corresponds to the reliable numerical scheme for a PDE derived in [12], where 3D linked and knotted scroll waves have been obtained, however, only for a short time interval and their stability is not analyzed.  Similarly, $N_P=19$ in Eq.(1) can also be considered as local coupling.  If so, we conclude that single-link chain chimera exist not only for the non-local but also for local coupling, and we expect that this can also be a robust pattern  for the respective PDE in the limit $N\rightarrow\infty, r\rightarrow 0$.  This situation is different from the Hopf link and trefoil stability, which have the pure non-local coupling origination (Ch.3); on the other hand, is similar to the rolls chimeras (scroll rings on $T^3$ ) , see Ch.2.
The double-link chain chimera exists in a smaller parameter region, see Fig. 10, which is detached from the locally coupled case.  Stability of the state begins with the non-locally coupling when $N_P=27$ ($r=0.01$ at $N=200$).  In our simulations, we have also tried to ''clone'' the chain chimeras  with three and more links. However, they appear to be unstable and are rapidly destroyed in the presence of even very small perturbations.

% \vspace*{-0.5cm} 
\subsection{Large-scale dynamics and transformations}

\hspace*{0.5cm} Our simulations show that multiple Hopf link and trefoil chimera states often change their structure with time in the following way.  When starting with 8 practically identical patterns placed in the unit cube, we observe soon afterwards  that they group into 4 visibly different  pairs, where the pairwise objects are only slightly different.  To our surprise, without or with only tiny perturbations ($10^{-4}$) the pairwise identity appears to be so strong that it practically is not affected by the asymmetry of the Runge-Kutta numerical algorithm, and  only imposed asymmetry in the initial conditions can slowly destroy it.  On the other hand, each such pair in the chimera state  has its own shape, size and position in the 3D space, distinct from the others.  This is illustrated by the video at http://chimera3d.biomed.kiev.ua/multiheaded/EvolutionT. 

Let us follow the dynamics of a multiheaded state inside the stability region shown Fig.\ref{fig:8}. Take the 8-headed trefoil or Hopf links obtained from the single trefoil but with stronger perturbations of the initial conditions and start the simulations.   Then, it is usually observed that in the some instants the samples collide and disappear or transform into other states.  Eventually, as a rule,  a hybrid state or a single-headed chimera is obtained. 

Typical evolutions of a 8-headed trefoil and Hopf links chimera states with rather strong perturbations ($0.1$) of the identical initial conditions are demonstrated in the video at http://chimera3d.biomed.kiev.ua/multiheaded/EvolutionS . As one can observe there, the dynamics result finally in 
a single trefoil, Hopf link or other kind of chimera states. It depends on the chosen initial cloning chimera, parameter values $\alpha$, $r$  and  perturbation value. Usually the trajectory evolutions were calculated  up to $t= 10^{4}$ for $N=200$.

To obtained the odd-headed Hopf links and trefoils chimera, an odd number of initial chimeras should be taken in the cloning procedure and placed in the unit cube. Sometimes, as we have often seen in the simulations, odd-headed chimeras arise from strong enough perturbations of the even-headed ones. In all cases considered, the probability to obtain a desired odd-headed chimera was rather small, however after some number of trials, we could eventually catch it.

Twice repeating the 8-cloning procedure gives birth to 64-headed Hopf links and trefoils.  As it is illustrated in Fig.\ref {fig:6}(i) and Fig.\ref{fig:7}(i), each such pattern consist of 16 groups per 4 similar elements inside. Moreover, each 4 groups among the  16 are quite similar too.  We expect that the head-adding sequence of the chimera states can be continued further, creating states with 512 and more objects.  It requires, clearly,  much more computational power to ensure the necessary accuracy of integrations.  Indeed, the  single Hopf link and trefoil states were obtained in our simulation of the $N^3$-dimensional network with $N=100$;  for the 8- and 64-headed states we had to take $N=200$  and  $N=400$, respectively.  In the latter case, a 64 million-dimensional nonlinear system should be integrated. The system complexity grows exponentially with further steps in the multiple chimera cascade.

\subsection{From trefoil to Hopf link}
\hspace*{0.5cm} In our simulations, we often observed the situation when a trefoil chimera state transforms into a Hopf link  (but not vise versa), see also~\cite{fm2000,ld2016}.  The transition starts in the moment when trefoil branches touch each other.  Soon afterwards the state transforms into a Hopf links or breaks down completely and disappears.

\begin{figure}[h]
\begin{center}
\resizebox{1.0\columnwidth}{!}{\includegraphics{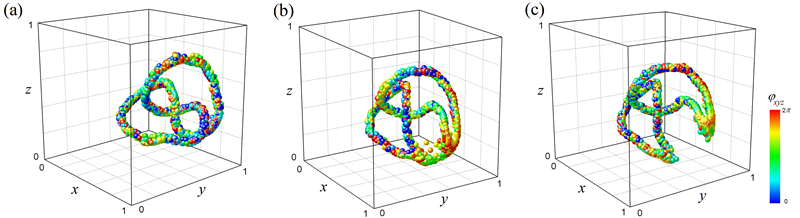}}
\end{center}
\caption{Trefoil - Hopf link transformation: (a) - trefoil ($t=900$), (b) -  transformation beginning ($t=990$), (c) - Hopf link ($t=1000$), $\alpha=0.632, r=0.04$,  $N = 100$. } 
\label{fig:11}
\end{figure}

There are three ways for the trefoil$\rightarrow$Hopf link transformation in Eq.(1). In the first case, when starting from random initial conditions, a trefoil is born,  exists for some time interval, and then transforms into a  Hopf link,  see illustrative video at http://chimera3d.biomed.kiev.ua/multiheaded/Hopflink. 

In the second case the trefoil$\rightarrow$Hopf link transformation occurs as a result of a strong enough perturbation of the initial conditions (generated originally a trefoil):
 http://chimera3d.biomed.kiev.ua/multiheaded/evolution.
  The  third  transformation scenario consists in the following:
take a single trefoil  inside its stability region, see Fig. 8, and move the  parameter point to the boundary. When close to the boundary,  it is often observed that trefoil can suddenly transform into a Hopf link. 

To illustrate the latter scenario, fix parameters $\alpha=0.6325, r=0.04$ close to the boundary of the trefoil stability region.
Shift the parameter $\alpha$ to $0.632$ and start to simulations. At $t=900$ the trefoil still exists (Fig. \ref{fig:11}(a))  but soon after, at $t=990$ two trefoil branches touch each other, and the transformation to a Hopf link begins ( in Fig. \ref{fig:11}(b)).  At  $t=1000$ (Fig. \ref{fig:11}(c)) a Hopf link is created, and it exists for long further simulations.

\section{Hybrid scroll wave chimeras}

\hspace*{0.5cm} In the previous sections,  different multiheaded scroll wave chimera states were reported for Eq.(1), each including similar 
incoherent elements such as rolls, Hopf links, trefoils,  or chains only.   Here,  we demonstrate a possibility of  hybrid-type organization for the multiheaded chimera states which can combine different of the above mentioned single-headed chimera types.

To obtain hybrid-type scroll wave chimeras we have tested different sample combinations in the cloning procedure.    There is no guarantee that the process will be successful. In many trials the cloned hybrid-type chimeras get destroyed very fast, and the procedure has to be repeated starting from a different number organization  of the initial 'chimera clones', also  varying the parameters and the magnitude of the perturbations.

\begin{figure}[ht!]
  \center{\includegraphics[width=1\linewidth]{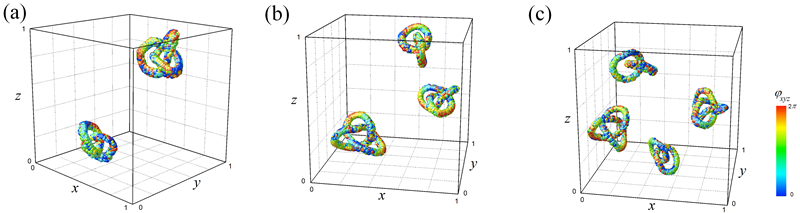}}
  \center{\includegraphics[width=1\linewidth]{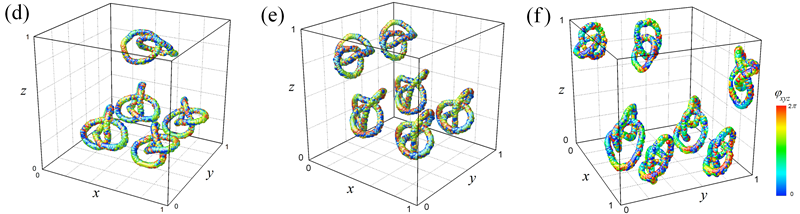}}
  \center{\includegraphics[width=1\linewidth]{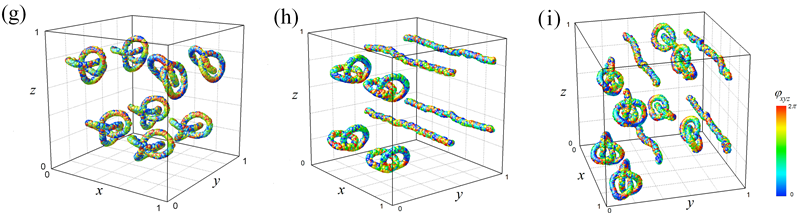}}
 \caption{Examples of multiheaded hybrid chimera states: 
(a) -  one trefoil and one Hopf links ($\alpha=0.785, r=0.02$), (b) - 2 trefoils and one Hopf links ($\alpha=0.755, r=0.02$), 
(c) -  2 trefoils and 2 Hopf links ($\alpha=0.735, r=0.02$),
 (d) - 4 trefoils and one Hopf links ($\alpha=0.625, r=0.02$),
(e) - 4 trefoils and 2 Hopf links ($\alpha=0.615, r=0.02$),
 (f) -  5 trefoils and 2 Hopf links ($\alpha=0.68, r=0.02$),
 (g) -  6 trefoils and 2 Hopf links ($\alpha=0.64, r=0.02$),
(h) - 4 Hopf links and 4 scroll wave rolls ($\alpha=0.7, r=0.02$).
(i) - 4 Hopf links, 4 trefoils and 4 scroll wave rolls ($\alpha=0.76, r=0.02$). $N = 200$. } 
  \label{fig:12}
\end{figure}

Screenshots of characteristic hybrid-type scroll wave  chimera states with additively growing number of samples are presented in Fig.\ref{fig:12} such as:   
(a) - Hopf link and trefoil,  (b) - Hopf link and 2 trefoils,  (c) - 2 Hopf links and 2 trefoils,   (d) -  Hopf link and 4 trefoils,  (e) - 2 Hopf links and 4 trefoils, 
 (f) -  2 Hopf links and 5 trefoils, 
 (g) -  2 Hopf links and 6 trefoils, 
(h) - 4 Hopf links and 4 rolls ($\alpha=0.7, r=0.02$),
(i) - 4 Hopf links, 4 trefoils and 4 rolls. 

These states exist for long times, up to $t= 10^{4}$ at least.  They are not stationary objects in 3D and are usually  characterized by the non-trivial  temporal large-scale dynamics. In our simulations we have also observed some other hybrid patterns, also preserving for long-time simulations.

 See videos at http://chimera3d.biomed.kiev.ua/multiheaded/hybrid, 
where more examples of the hybrid-type chimeras are shown. 

\begin{figure}[h]
\center{\resizebox{0.65\columnwidth}{!}{\includegraphics{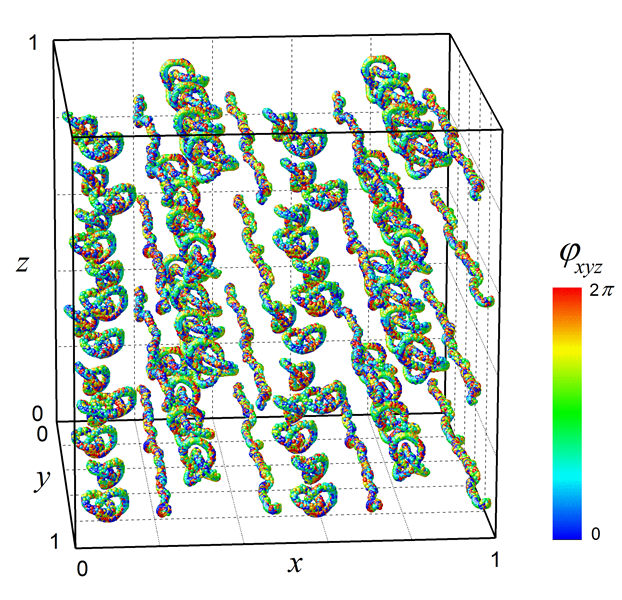}}}
 \caption{Example of 80-headed hybrid chimera states consisting of 32 trefoils, 32  Hopf links, and 16 rolls  ($\alpha=0.76, r=0.01$). $N = 400$. } 
  \label{fig:13}
\end{figure}

To finalize, our last example is a 80-headed scroll wave chimera state including  32 Hopf links, 32 trefoils, and 16 rolls illustrated in
Fig.\ref{fig:13}. The state is calculated for N=400 (i.e. 64 million oscillators), simulation time was $t=1000$.

%The decreasing parameters $r$ and $\alpha$ leads to growing the trefoil size moreover, its incoherent filaments become thicker. In the case, however, the result is similar: trefoil will eventually gives rise to the Hopf links or  
%may disappear dropping down to the regular oscillations.  Under some other parameter variations, two neighboring trefoils or Hopf links can touch each other producing sometimes the parallel rolled chimera (as in Ch.II).   

\vspace*{-0.5cm}
\section{Conclusion} 

\begin{figure}[h]\sidecaption
\resizebox{0.61\columnwidth}{!}{\includegraphics{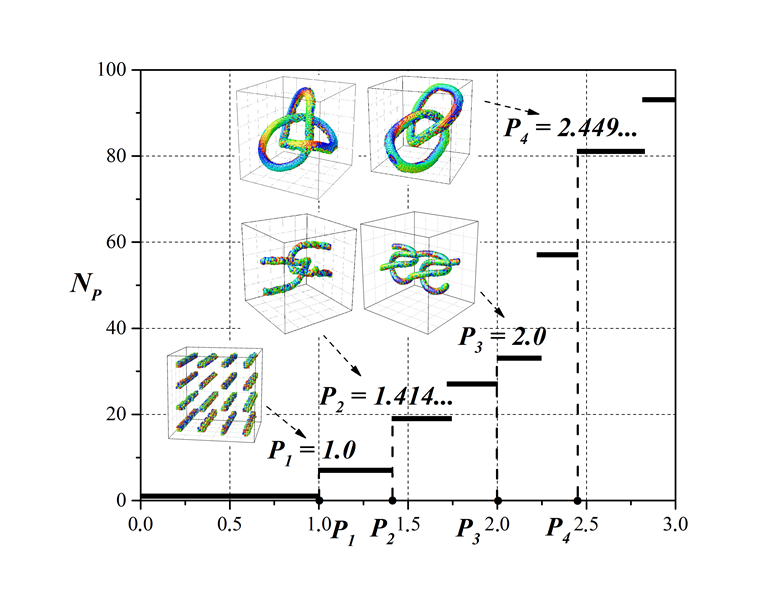}}
\caption{
Number of oscillators $N_P$ coupled to each oscillator in the model (1) depending on the value of $P=rN$. Parallel and crossed rolled chimeras exist beginning from $P=P_{1}=1.0$ ($N_{P}=7$), single-link chains - from $P=P_{2} =  1.414...$ ($N_{P}=19$), double-link  chains - from $P=P_{3} =  2.0$ ($N_{P}=27$), trefoils and Hopf links - from $P=P_{4} =  2.44948...$ ($N_{P}=81$). The minimal values  $N_{P}$ are found to be the same  for $N=100, 200$, and $400$. }
\label{fig:14}
\end{figure}

\hspace*{0.5cm} We have demonstrated a diversity of multiple scroll wave chimeras for the  three-dimensional network of coupled Kuramoto phase oscillators with non-local coupling.  Wide parameter regions are obtained for rolls, chains, Hopf links, and trefoil patterns.  It follows, in particular, that rolled chimeras exist not only for non-local but also for local coupling schemes, beginning from only $N_P=7$ nearest-neighbor couplers (as in the simplest diffusive coupling scheme). This fact is schematically indicated by the left corner inset in Fig. \ref{fig:14}. 

The next chimera state to appear when increasing the coupling radius $r=N/P$ in Eq.~(1) is the single-link chain, this occurs in the case of  $N_P=19$ couplers. At further increase of the coupling radius $r$, first, the double-link chain stabilizes for $N_P=27$ couplers ($r=0.01$ at $N=200$). 
Then, only at $N_P=81$  ($r=0.02449$) a variety in single and multiple Hopf links and trefoils become stable to exist further for this and larger values of $r$, up to appr. 0.08.

An essential condition for the scroll wave chimeras appearance is the  intermediate value of the phase shift $\alpha$, appr. between 0.6 and 0.9.  Therefore, Hopf link and trefoil chimera states exist in the Kuramoto model only with essentially non-local coupling and sufficiently large phase shift. On the other hand, multiple rolled chimeras  which are actually the scroll rings in the circular coordinates are more stable objects. They grow not only for non-local but even for local,  diffusive-type coupling schemes.  Single-link chain chimeras are also preserved in the locally coupling case, but not the double-link ones which require some level of non-locality for the stabilization.  We believe that the described  fascinating scroll wave chimeras can be found in other, more realistic 3D networks displaying one of the inherent features of nature, that is due to non-local coupling.  

\section*{Acknowledgments} 
\hspace*{0.5cm}
\noindent We thank  B.Fiedler,  E.Knobloch, P.Manneville  and M.Hasler for
illuminating discussions, and 
the Ukrainian Grid Infrastructure for providing the computing
cluster resources and the parallel and distributed software.

%\section*{References}

\end{document}